\documentclass{article}
\usepackage[utf8]{inputenc}

\usepackage{amsmath,amssymb,bbm,amsthm}
\usepackage{fullpage}
\usepackage{thm-restate,color,xcolor,xspace}
\usepackage{hyperref,cleveref}
\usepackage{algorithm}
\PassOptionsToPackage{noend}{algpseudocode}
\usepackage{algpseudocode}
\usepackage{thm-restate}
\usepackage{graphicx}
\usepackage{tikz}
\usepackage{colortbl}

\algrenewcommand\algorithmicrequire{\textbf{Input:}}

\usepackage{amsmath}

\newtheorem{theorem}{Theorem}[section]
\newtheorem{lemma}[theorem]{Lemma}

\newtheorem*{theorem*}{Theorem}
\newtheorem{definition}[theorem]{Definition}
\newtheorem{corollary}[theorem]{Corollary}
\newtheorem{remark}[theorem]{Remark}

\newtheorem{observation}[theorem]{Observation}
\newtheorem{claim}[theorem]{Claim}

\newcommand{\E}{\mathsf{E}}

\newcommand{\var}{\mathsf{Var}}

\definecolor{verylightgray}{gray}{0.9}

\title{Optimality of Frequency Moment Estimation}
\author{
  Mark Braverman\\
  \emph{Princeton University}
  \and
  Or Zamir\\
  \emph{Tel Aviv University}
}
\date{}
\begin{document}

\maketitle

\begin{abstract}
    Estimating the second frequency moment of a stream up to~$(1\pm\varepsilon)$ multiplicative error requires at most~$O(\log n / \varepsilon^2)$ bits of space, due to a seminal result of Alon, Matias, and Szegedy.
    It is also known that at least~$\Omega(\log n + 1/\varepsilon^{2})$ space is needed.
    We prove a tight lower bound of~$\Omega\left(\log \left(n \varepsilon^2 \right) / \varepsilon^2\right)$ for all~$\varepsilon = \Omega(1/\sqrt{n})$. 
    Note that when~$\varepsilon>n^{-1/2 + c}$, where~$c>0$, our lower bound matches the classic upper bound of AMS. For smaller values of~$\varepsilon$ we also introduce a revised algorithm that improves the classic AMS bound and matches our lower bound.
\end{abstract}

\section{Introduction}
An extensive body of literature is devoted to the streaming model of computation, which is important for the analysis of massive datasets and in network traffic monitoring.
A central problem in this model is the \emph{frequency moment estimation} problem:
Elements from a universe~$U$ are given to the algorithm one-by-one, defining a vector of frequencies --- that is,~$f_x \in \mathbb{N}$ is the number of times the element~$x\in U$ appeared in the stream; 
Finally, the algorithm has to return, with good probability, a~$(1\pm \varepsilon)$-estimation of~$F_p := \sum_{x\in U} f_x^p$ --- the \emph{$p$-th frequency moment} of the stream. We generally denote the length of the stream by~$n$ and assume that~$|U|=\text{poly} (n)$. The main complexity parameter studied in this model is how much \emph{space} is needed for the algorithm to succeed. 
The study of both the streaming model and of frequency moment estimation in it was initiated in the seminal 1996 work of Alon, Matias, and Szegedy \cite{alon1996space}.

The case of~$p=2$, or \emph{second moment estimation}, is of particular importance. It is often called the \emph{repeat rate} or \emph{surprise index}, and is used in various tasks such as database query optimization~\cite{alon1999tracking}, network traffic anomaly detection~\cite{krishnamurthy2003sketch}, approximate histogram maintenance~\cite{gilbert2002fast} and more.
Other moments of particular interest are~$p=1$, corresponding to the \emph{approximate counting} problem~\cite{morris1978counting,nelson2022optimal}, and~$p=0$, corresponding to the \emph{distinct elements} problem~\cite{flajolet1985probabilistic,indyk2003tight,kane2010optimal}.
Among these special cases, only the space complexity of the first remains not fully understood. 
The original algorithm for~$F_2$-estimation given by Alon, Matias, and Szegedy  uses~$O(\log n / \varepsilon^2)$ bits of space; while the highest known lower bound due to Woodruff in 2004~\cite{woodruff2004optimal} is~$\Omega(\log n + 1/\varepsilon^2)$ --- leaving up to a quadratic gap between the upper and lower bounds for certain choices of~$\varepsilon$.

While~$F_p$-estimation for~$p\leq 2$ uses amount of space that is only logarithmic in the length of the stream, it was shown that for~$p>2$ at least~$\Omega(n^{1-2/p} / \text{poly}(\varepsilon))$ space is needed~\cite{bar2004information,chakrabarti2003near} --- which is polynomial in the stream's length. 
A long list of works~\cite{indyk2005optimal,bhuvanagiri2006simpler,monemizadeh20101,andoni2011streaming,braverman2010recursive,andoni2017high,ganguly2011polynomial,woodruff2012tight,ganguly2011lower,li2013tight} resulted in a nearly-tight bound of~$\tilde{\Theta}\left(n^{1-2/p}/\varepsilon^2\right)$ for~$F_p$-estimation for every~$p>2$ (not necessarily an integer) and~$\varepsilon$, for some ranges of parameters the bounds are tight --- in others there is a gap between the bounds that is poly-logarithmic in the bound itself. 

For~$p\leq 2$ the space complexity is not as well understood.
Woodruff~\cite{woodruff2004optimal} showed a lower bound of~$\Omega(\log n + 1/\varepsilon^2)$ for every~$p\neq 1$, this is optimal in terms of~$\varepsilon$ alone and is also known to be optimal for the distinct elements problem (that is, $p=0$).
For the special case of approximate counting (that is, $p=1$), a tight bound of~$\Theta(\log\log n + \log \varepsilon^{-1})$ is known~\cite{nelson2022optimal}.
For the range of~$p\in [0,1)$, the upper bound of AMS was improved by Jayaram and Woodruff who presented a nearly-tight~$\tilde{O}\left(\log n + 1/\varepsilon^2\right)$ bound in that range~\cite{jayaram2019towards}.
This leaves~$p\in (1,2]$ as the last remaining range within no nearly-tight bounds are known. 

For certain generalizations more is known: 
When the stream is \emph{randomly shuffled} and given in random order, then in the range~$p\in (1,2)$ (excluding~$p=2$) \cite{braverman2018revisiting} showed an improved upper bound of~$\tilde{O}\left(\log n + 1/\varepsilon^2\right)$.
When \emph{updates} are allowed in the stream, that is, elements can also be deleted and not only added to it, then \cite{kane2010exact} showed that $\Theta(\log n / \varepsilon^2)$ is optimal for~$p\leq 2$.

In this work, we settle the space complexity of second frequency moment estimation, the only remaining integral frequency moment for which the space complexity is not understood. 
For~$F_2$, we show that the AMS bound is essentially tight.
\begin{theorem*}
    Let~$\mathcal{A}$ be a streaming algorithm that gives an~$\left(1\pm \varepsilon\right)$  multiplicative approximation to the~$F_2$ of its input stream and succeeds with probability~$\geq 2/3$, for some~$\varepsilon=\Omega(1/\sqrt{n})$.
    Then, the space used by~$\mathcal{A}$ is~$\Omega\left(\log\left(\varepsilon^2 n\right)/\varepsilon^2\right)$.
\end{theorem*}

Note that the range~$\varepsilon<1/\sqrt{n}$ is less interesting as~$O\left(n\log\left(1+\frac{|U|}{n}\right)\right)$ space suffices for exactly maintaining the vector of frequencies.\footnote{In fact, we may always assume that~$|U|=O(n^2)$ as a hash mapping to this size is likely to not cause any collisions. Thus,~$O(n\log n)$ space always suffices to maintain an exact histogram of the stream.}
We observe that in the range where~$\varepsilon$ is very close to~$1/\sqrt{n}$ our lower bound is (slightly) lower than the AMS upper bound, we show that this is inherent by introducing a modification of the AMS algorithm that matches our lower bound in this range.

\begin{theorem*}
    For~$\varepsilon=\Omega(1/\sqrt{n})$, we can get a~$(1\pm \varepsilon)$-approximation of the~$F_2$ of a stream of length~$n$ using~$O\left(\log\left(\varepsilon^2 n\right)/\varepsilon^2\right)$ space with success probability~$>2/3$.
\end{theorem*}

Most of the lower bounds for streaming problems are based on reductions from \emph{communication complexity}. In~\cite{jayaram2019towards}, a natural barrier to prove a better than~$\tilde{\Omega}(1/\varepsilon^2)$ lower bound was shown: Even in a very strong model of communication,~$O\left(1/\varepsilon^2 \cdot \left(\log\log n + \log d + \log\varepsilon^{-1}\right)\right)$ bits of communication suffice for the players to correctly produce a~$F_p$ estimation, where~$d$ is the diameter of the communication graph. 
This means that problems who reduce to~$F_p$-estimation have a too low communication complexity to improve the existing lower bounds.
To overcome this natural barrier, we present a new type of a \emph{direct sum theorem} that takes place at the level of the streaming algorithm rather than the level of the communication model --- informally, we pack many instances of problems with communication complexity~$\Theta(1/\varepsilon^2)$ into a single stream, and then directly show that a successful streaming algorithm must solve them all. 
In Section~\ref{sec:overview} we give a detailed high-level overview of our proofs.
The lower bound is presented in Section~\ref{sec:EDISJ} and Section~\ref{sec:lower}. The improved upper bound is presented in Section~\ref{sec:upper}.
We conclude and present remaining open problems in Section~\ref{sec:open}.

\paragraph{Erratum:}
An earlier version of this paper claimed that our lower bound for~$F_2$ extends to all non-integral frequency moments with~$p \in (1,2]$, and included a brief high-level sketch of this generalization. It has since been brought to our attention that the proposed argument does not establish the claimed bound for~$p \in (1,2)$. In particular, the sketch outlined how to set parameters so that solving the~$F_p$ estimation problem would yield solutions to all instances of the Exam Disjointness problem and their direct sum, but did not address the complexity of Disjointness with these parameters. The straightforward completion of this detail does not yield the desired lower bound. We therefore retract this claim and restrict our result to the case~$p = 2$.

\section{High-Level Overview}\label{sec:overview}
In this Section we give a high-level overview of the components used in our lower and upper bounds. The lower bound is then proven in Section~\ref{sec:EDISJ} and Section~\ref{sec:lower}, and the upper bound in Section~\ref{sec:upper}.

\subsection{Exam Communication Model}
As in many other streaming lower bounds, our initial building blocks are reductions from communication complexity problems. However, instead of using the classic communication model, we consider a slight variant which we call \emph{the exam model}.
In the classic numbers-in-hand communication model,~$t$-players receive different parts~$x_1,x_2,\ldots,x_t$ of an input and have to compute some function~$f(x_1,\ldots,x_n)$ of the entire input by communicating with each other. 
In our model, we introduce an additional player, \emph{a referee}, who has an additional input~$y$ which we think of as a \emph{question} about the players' inputs~$\{x_i\}$. 
The players still receive their parts of the input and are allowed to communicate with each other, but not with the referee; finally, a player sends a single message to the referee, and the referee then has to compute some function~$g(x_1,\ldots,x_t,y)$ of \emph{both} the input and her ``secret" question~$y$.

In the classic \emph{Disjointness} problem, each of the~$t$ players receives a set~$S_i$ and they have to determine whether or not all of their input sets are pairwise disjoint. It is known that solving this problem remains hard even under the promise that the input sets are either all disjoint or contain a unique element appearing in all input sets and do not intersect further.
We introduce a variant we call \emph{Exam Disjointness} in which each of the players still receives a set~$S_i$, and the referee receives a single element~$y$ from the universe; based on the message she receives the referee needs to decide whether~$y$ appears in \emph{all} input sets~$S_i$. In other words, instead of determining whether there was a unique intersection between their input sets, the players now have to send enough information to determine if a specific~$y$ given to the referee is that unique intersection.
In Section~\ref{sec:EDISJ} we prove that solving Exam Disjointness requires at least as much information about the input as the standard Disjointness problem.

This communication model turns out to be very useful for reducing a communication problem to a one-pass streaming algorithm problem. Intuitively, only the suffix a of the stream would depend on the referee's input~$y$, which would in turn mean that the algorithm processing the rest of the stream has to learn enough information to answer any possible question of the referee that might appear in the exam.
In Section~\ref{subsec:EDISJtoF2}, we use the Exam Disjointness problem to recover the existing~$\Omega(1/\varepsilon^2)$ lower bound for~$F_2$-estimation, with a simple and short proof.

\subsection{Direct Sum for Dependent Instances}
A standard tool for proving lower bounds is proving \emph{a direct sum theorem}; this means showing that solving several \emph{independent} instances of a certain problem is as hard as solving every one of them separately.
In our proof, we introduce an unintutive variant of a direct sum theorem in which several instances \emph{are} dependent, and nevertheless solving them all is as hard as solving them separately.

\begin{figure}
    \centering
    
    \begin{tikzpicture}
    \def\cellwidth{0.5}
    \def\cellheight{0.5}

    \path[use as bounding box] (-0.5*\cellwidth, -0.3) rectangle (31*\cellwidth, \cellheight);

    \node[left] at (-0.5*\cellwidth, 0.5*\cellheight) {$\ell=1$};

    \def\cellwidth{2}
    \foreach \i in {0,...,7} {
        \pgfmathsetmacro{\x}{\i * \cellwidth}
        
        \pgfmathparse{int(mod(\i, 4))}
        \let\modresult\pgfmathresult

        \ifnum \modresult<1
            \fill[gray] (\x,0) rectangle ++(\cellwidth, \cellheight);
        \else
            \fill[white] (\x,0) rectangle ++(\cellwidth, \cellheight);
        \fi
        
        \draw[black] (\x,0) rectangle ++(\cellwidth, \cellheight);
        
        \node at (\x + 0.5*\cellwidth, 0.5*\cellheight) {};
    }
    \end{tikzpicture}

    \begin{tikzpicture}
    \def\cellwidth{0.5}
    \def\cellheight{0.5}

    \path[use as bounding box] (-0.5*\cellwidth, -0.3) rectangle (31*\cellwidth, \cellheight);

    \node[left] at (-0.5*\cellwidth, 0.5*\cellheight) {$\ell=2$};

    \def\cellwidth{1}
    
    \foreach \i in {0,...,15} {
        \pgfmathsetmacro{\x}{\i * \cellwidth}
        
        \pgfmathparse{int(mod(\i, 4))}
        \let\modresult\pgfmathresult

        \ifnum \modresult<1
            \fill[gray] (\x,0) rectangle ++(\cellwidth, \cellheight);
        \else
            \fill[white] (\x,0) rectangle ++(\cellwidth, \cellheight);
        \fi
        
        \draw[black] (\x,0) rectangle ++(\cellwidth, \cellheight);
        
        \node at (\x + 0.5*\cellwidth, 0.5*\cellheight) {};
    }
    \end{tikzpicture}
    
    \begin{tikzpicture}
    \def\cellwidth{0.5}
    \def\cellheight{0.5}

    \path[use as bounding box] (-0.5*\cellwidth, -0.3) rectangle (31*\cellwidth, \cellheight);

    \node[left] at (-0.5*\cellwidth, 0.5*\cellheight) {$\ell=3$};

    \foreach \i in {0,...,31} {
        \pgfmathsetmacro{\x}{\i * \cellwidth}
        
        \pgfmathparse{int(mod(\i, 4))}
        \let\modresult\pgfmathresult

        \ifnum \modresult<1
            \fill[gray] (\x,0) rectangle ++(\cellwidth, \cellheight);
        \else
            \fill[white] (\x,0) rectangle ++(\cellwidth, \cellheight);
        \fi
        
        \draw[black] (\x,0) rectangle ++(\cellwidth, \cellheight);
        
        \node at (\x + 0.5*\cellwidth, 0.5*\cellheight) {};
    }
    \end{tikzpicture}

    \caption{The parts of the stream corresponding to the players of a $2^\ell$-party Exam Disjointness instance, for~$\ell=1,2,3$.}
    \label{fig:several_instances}
\end{figure}
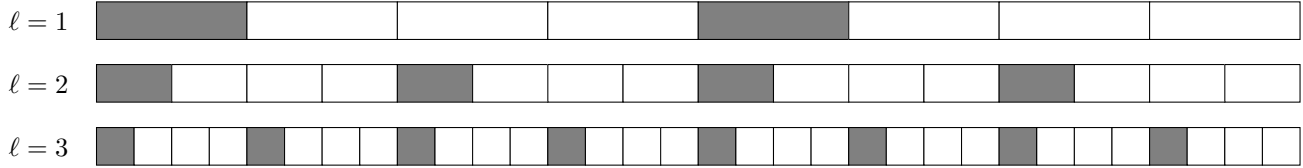

As stated above, we first show that we can reduce the~$t$-party Exam Disjointness problem to estimating the~$F_2$ of a certain stream. We then introduce a distribution over input stream that encodes within it instances of~$t$-party Exam Disjointness for multiple different values of~$t$. See Figure~\ref{fig:several_instances} for an example of a stream in which Exam Disjointness instances with two, four, and eight players are encoded, using different parts of the stream to encode each player; while at the same level all players get a disjoint part of the stream --- players at different levels occasionally intersect.
We would describe a construction containing roughly~$\log \left(\varepsilon^2 n\right)$ such levels, such that solving the Exam Disjointness instance corresponding to a single level requires~$\Omega(1/\varepsilon^2)$ bits of space. 
If instances of different levels were disjoint, a direct sum theorem would yield our desired~$\Omega\left(\log \left(\varepsilon^2 n\right)/\varepsilon^2\right)$ lower bound; we obtain a similar bound despite the intersection of different levels.

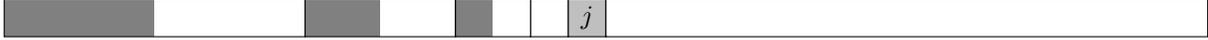
\begin{figure}
    \centering
    
\begin{tikzpicture}
    \def\cellwidth{0.5}
    \def\cellheight{0.5}

    \fill[gray] (0, 0) rectangle ++(4*\cellwidth, \cellheight); 
    \fill[white] (4*\cellwidth, 0) rectangle ++(4*\cellwidth, \cellheight); 
    \draw[black] (0, 0) rectangle ++(8*\cellwidth, \cellheight);
    \node at (3.5*\cellwidth, 0.5*\cellheight) {};

    \fill[gray] (8*\cellwidth, 0) rectangle ++(2*\cellwidth, \cellheight); 
    \fill[white] (10*\cellwidth, 0) rectangle ++(2*\cellwidth, \cellheight); 
    \draw[black] (8*\cellwidth, 0) rectangle ++(4*\cellwidth, \cellheight);
    \node at (10*\cellwidth, 0.5*\cellheight) {};

    \fill[gray] (12*\cellwidth, 0) rectangle ++(\cellwidth, \cellheight); 
    \fill[white] (13*\cellwidth, 0) rectangle ++(\cellwidth, \cellheight); 
    \draw[black] (12*\cellwidth, 0) rectangle ++(2*\cellwidth, \cellheight);
    \node at (12.5*\cellwidth, 0.5*\cellheight) {};

    \fill[white] (14*\cellwidth, 0) rectangle ++(\cellwidth, \cellheight);
    \draw[black] (14*\cellwidth, 0) rectangle ++(\cellwidth, \cellheight);
    \node at (14.5*\cellwidth, 0.5*\cellheight) {};

    \fill[lightgray] (15*\cellwidth, 0) rectangle ++(\cellwidth, \cellheight);
    \draw[black] (15*\cellwidth, 0) rectangle ++(\cellwidth, \cellheight);
    \node at (15.5*\cellwidth, 0.5*\cellheight) {$j$};

    \fill[white] (16*\cellwidth, 0) rectangle ++(16*\cellwidth, \cellheight);
    \draw[black] (16*\cellwidth, 0) rectangle ++(16*\cellwidth, \cellheight);
    \node at (16*\cellwidth, 0.5*\cellheight) {};

\end{tikzpicture}

    \caption{An index~$j$ and its preceding players from several different levels.}
    \label{fig:preceding_players}
\end{figure}

The core idea of this part is the following observation:
Consider an average index~$j$ within the stream; at each level, look at the part of the stream that encodes the last player appearing before index~$j$ in that level.
Although some of these players might intersect each other, usually there is a large subset of them who are pairwise disjoint. See Figure~\ref{fig:preceding_players} for an illustration. At index~$j$ we must have information about each level, as otherwise we would not be able to finally solve the instance of that level.  For the levels in which the last players are pairwise disjoint, we would be able to show that the information known about each of them by~$j$ is independent. This would be enough to prove a ``local" direct sum for only a subset of the levels at each index~$j$, and an average over all indices would then give us a bound as good as a ``global" direct sum.
The exact details are given in Section~\ref{subsec:manyEDISJ}.

Another way to look at the direct sum across levels is that at each level $\ell$ we get a lower bound on the amount of information the stream typically needs to store about the $\sim 2^\ell$ preceding stream elements. It is then shown that these lower bounds add up, as these different-scaled ``pasts" are essentially disjoint. This multi-scale phenomenon is arguably the ``real reason" behind the $\log n$ multiplicative factor in the lower bound. A similar phenomenon was used to show that estimating the majority of $n$ random bits requires $\Omega(\log n)$ memory \cite{braverman2020coin}.

\subsection{Improved Algorithm for The Small Error Regime}
The classic algorithm of AMS gives us a bound of~$O(\log n / \varepsilon^2)$ for~$(1\pm\varepsilon)$-estimating the~$F_2$ of a stream.
Evidently, this matches our~$\Omega\left(\log \left(\varepsilon^2 n\right)/\varepsilon^2\right)$ lower bound only when~$\varepsilon$ is polynomially larger than~$n^{-1/2}$.
In Section~\ref{sec:upper} we introduce a slight variant of the AMS algorithm that matches our lower bound. 
The modification is rather simple: we randomly partition the universe of elements into roughly~$1/\varepsilon^2$ disjoint subsets, and then run the AMS algorithm separately (and simultaneously) on the subsets of the stream contained in each of these parts. Each subset of the stream only contains around~$\varepsilon^2 n$ elements, which would result in the improved bound.
A similar modification was used before in algorithms aiming to reduce the memory-probes-per-update complexity of the AMS algorithm~\cite{thorup2004tabulation}.

\section{Mutual Information and Streaming Algorithms}
We frequently use the notions of information complexity, mutual information and entropy, see~\cite{bar2004information} for example for more background on these notions.

A \emph{one-pass} streaming algorithm receives a sequence of inputs~$X=\left(X_1,\ldots,X_n\right)$ one by one, and eventually outputs an answer. Denote by~$M_1,\ldots,M_n$ the \emph{memory transcript} of the streaming algorithm. That is,~$M_i$ is the memory state of the algorithm immediately after receiving the~$i$-th input~$X_i$.
We observe that~$M_i$ is drawn from a distribution that depends solely on~$M_{i-1}$ and~$X_i$.
From now on, we think of the inputs as a distribution, thus every~$X_i$ and every~$M_i$ is a random variable.

The following classic observation highlights the benefit of studying the mutual information between the input distribution and the memory transcript.
\begin{observation}\label{obs:memoryfrominfo}
    The number of memory bits used by a streaming algorithm is at least \[
    \max_j \left(I\left(M_j \ ; \ X\right)\right)
    .
    \]
\end{observation}
\begin{proof}
    It follows immediately as~$I(M_j \ ; \ X)\leq H(M_j) \leq \log |\text{Support}\left(M_j\right)|$. 
\end{proof}

We next introduce a useful lemma.

\begin{lemma}\label{lem:streaminginfo}
    If the inputs~$X_i$ are mutually independent, then
    \[
    I(M \ ; \ X) = \sum_{i=1}^{n} I\left(X_i \ ; \ M_i \ | \ M_{i-1}\right)
    .\]
\end{lemma}
\begin{proof}
    Consider the following chain of equalities,
    \begin{align*}
        I(X_1,\ldots,X_n \ ; \ M_1,\ldots,M_n) &= \sum_{i=1}^{n} I\left(X_1,\ldots,X_n \ ; \ M_i \ | \ M_{<i}\right) \tag{1}\\
        &= \sum_{i=1}^{n} \left(
        I\left(X_i \ ; \ M_i \ | \ M_{<i}\right) +
        I\left(X \ ; \ M_i \ | \ M_{<i}, X_i\right) 
        \right)\tag{2}\\
        &= \sum_{i=1}^{n} \left(
        I\left(X_i \ ; \ M_i \ | \ M_{i-1}\right) +
        0 
        \right).\tag{3}
    \end{align*}
    Inequalities~$(1)$ and~$(2)$ are applications of the chain rule of mutual information.
    The second part of Inequality~$(3)$, that is~$I\left(X \ ; \ M_i \ | \ M_{<i}, X_i\right)=0$, follows as~$X$ is independent of the algorithm's random coins while~$M_i$ is fully determined by~$M_{<i},X_i$ and these random coins.
    The first part of Inequality~$(3)$, that is~$I\left(X_i \ ; \ M_i \ | \ M_{<i}\right) = I\left(X_i \ ; \ M_i \ | \ M_{i-1}\right)$, follows as to both~$X_i$ and~$M_i$ conditioning on~$M_{<i}$ or on~$M_{i-1}$ is equivalent, this uses the independence of~$X_i$ from~$X_{<i}$.
\end{proof}

\section{Exam Variant of Multi-Party Set Disjointness}\label{sec:EDISJ}
At the core of our lower bound is a variant of the Multi-Party Set Disjointness problem (denoted DISJ from now on).
The DISJ$_t$ communication problem is defined as follows. 
There are~$t$ players, each player~$i\in [t]$ receives a set $S_i$ of elements from some universe~$U$; We are promised that either all sets are disjoint, or otherwise there is exactly one element that appears in all of the sets and beside it all other elements are disjoint; the players need to decide in which of the two cases their input sets are. 
We focus on the \emph{one-way communication model} in which there is a predetermined order of the players; In its turn, a player can look at its own input set and on the communication it received from the previous player, work for an unbounded amount of time, and then send a communication to the next player; The last player needs to answer whether the sets were disjoint or contained a shared element. 
Chakrabarti et al.~\cite{1214414} proved that the total length (in bits) of the message communicated by the players throughout the protocol must be at least~$\Omega\left(\frac{1}{t} \sum_{i=1}^{t} |S_i| \right)$ if they succeed with good probability.
Intuitively, this follows as otherwise the total amount of communicated information is smaller than even a single player's set, and thus even a single set cannot be fully described throughout the protocol's transcript.
Their lower bound is optimal and holds even in a stronger communication model called \emph{the blackboard model} in which each player can see the messages communicated by all previous players, rather than just the preceding one.
The same lower bound holds even in the stronger model of communication in which players can talk several times in arbitrary order and not only in a one-way fashion \cite{gronemeier2009asymptotically, jayram2009hellinger}.

For our needs, we introduce a slight variant of the one-way DISJ$_t$ problem which we call \emph{Exam DISJ$_t$} and denote by EDISJ$_t$.
In this version, we still have~$t$ players with input sets~$S_i$ that can be disjoint or have a unique common intersection, but we also have an additional special player, which we call \emph{the referee}. 
The referee receives as input a single universe element~$x\in U$.
The~$t$ players still communicate one by one according to order, and the last player communicates a message to the referee. Then, the referee has to decide (and succeed with probability~$\geq 2/3$) whether there was an intersection between the players' sets \emph{and} that intersection is exactly the referee's input~$x$.
We think of this as an ``exam" for the players: If they claim there was an intersection, they also need to know enough about the intersection to tell whether or not it is~$x$.
In this section, we extend the lower bound for DISJ$_t$ and prove that EDISJ$_t$ also requires as much communication. 

Let~$\mathcal{P}$ be a protocol that solves EDISJ$_t$, and denote by~$\Pi = \Pi(X)$ its \emph{transcript} on an input~$X=\left(S_1,S_2,\ldots,S_t,x\right)$; the transcript consists of the message communicated by each of the~$t$ players during its turn to speak. Note that~$\Pi$ is a random variable that depends only on~$X':=(S_1,\ldots S_t)$, as the referee does not speak and thus~$x$ does not affect the transcript.

We define an input distribution for the players of EDISJ$_t$. Let~$m$ be a size parameter, and let~$U$ be a universe of size~$|U|\geq m^{4}$.
Each input set~$S_i$ is drawn randomly and independently as a subset of~$U$ of size~$\lceil \frac{m}{t} \rceil$.
We note that~$m\leq \sum_{i=1}^{t} |S_i| < m+t$ and that with high probability all~$t$ sets are disjoint. 
Denote this distribution of~$X'$ by~$\mu = \mu\left(t,m,U\right)$.

In this section, we prove the following lower bound on the information complexity of EDISJ$_t$ on the distribution~$\mu$. We also assume that~$t < \frac{m}{4 \ln m}$.
\begin{theorem}\label{thm:EDISJ}
    Let~$\mathcal{P}$ be a protocol that solves EDISJ$_t$ with success probability~$\geq 2/3$ for any input with~$|S_i|\leq \lceil \frac{m}{t} \rceil$ for~$i\in [t]$.
    Let~$X'=(S_1,\ldots,S_t)\sim \mu$ and denote by~$\Pi = \Pi(X')$ the transcript of~$\mathcal{P}$ on~$X'$.
    Then,
    \[
    I\left(X' \ ; \ \Pi\right) = \Omega\left(\frac{m}{t}\right)
    .
    \]
\end{theorem}

The proof of Theorem~\ref{thm:EDISJ} is similar to standard lower bound proof for multi-party set disjointness. It proceeds in two steps: first, derive an appropriate information bound for the AND function -- which is the one-bit version of disjointness; second, apply a direct sum argument to obtain the lower bound statement. 

\subsection{AND Lower Bounds}
All proofs for the hardness of the standard Disjointess problem go through a direct sum between instances of the AND problem~\cite{bar2004information,chakrabarti2003near,jayram2009hellinger}.
In the AND$^t$ communication problem, there are~$t$ players and each receives a single input bit~$Y_i \in \mathbb{F}_2$; the players need to compute the ``and" function~$\bigwedge_{i=1}^t Y_i$. Also denote by~$Y=\left(Y_1,\ldots,Y_t\right)$.
In the DISJ$_t$ problem, we may think of each player's input as the characteristic vector of its set~$S_i \subset U$, and denote its coordinates by~$Y_{i,u}=1$ if~$u\in S_i$ and otherwise~$0$.
Then, the answer to~DISJ$_t$ is simply~$\bigvee_{u\in U} \bigwedge_{i=1}^{t} Y_{i,u}$.
Due to this structure, techniques introduced by~\cite{bar2004information} show it suffices to give a lower bound to the information complexity of~AND$^t$ to imply a lower bound for~DISJ$_t$.
Consider the probability distribution~$\nu$ over inputs to the~AND$^t$ problem that gives probability~$\frac{1}{2}$ for all input bits being~$0$, and probability~$\frac{1}{2}$ for exactly one coordinate chosen at random being~$1$ and the rest~$0$.
Let~$\Pi$ be the transcript of a communication protocol that solves~AND$^t$ with probability~$>2/3$, the citations above show that $I\left( Y \ ; \ \Pi \right) = \Omega\left(\frac{1}{t}\right)$.
Denote by~$\Pi_t$ the last message communicated in the protocol~$\Pi$, from which the answer has to be deduced (in a one-way communication setting this is simply the message sent by the last player).
In slightly more detail, they show that as long as there is a non-negligible statistical difference between the distributions of~$\Pi_t$ when the input is~$0^t$ and when the input is~$1^t$ (which there must be for solving~AND$^t$ correctly), then the above information bound holds.

\begin{lemma}[e.g., Corollary 8 of \cite{jayram2009hellinger}\protect\footnote{In \cite{jayram2009hellinger} this statement is phrased in terms of the Hellinger distance between the two distributions rather than the total variation distance, but in the constant regime the two distances are equivalent; See for example Proposition A.2 in~\cite{bar2004information}.}]\label{lem:AND_lb}
    If $\|\Pi_t(1^t)-\Pi_t(0^t)\|_1>0.1$, then
    $I_{\nu}\left( Y \ ; \ \Pi \right) = \Omega\left(\frac{1}{t}\right)$.
\end{lemma}

\begin{remark}
Many formulations of this lower bound, such as in~\cite{jayram2009hellinger}, are stated in terms of the \emph{conditional} information cost~$I(Y; \Pi \mid G)$, where~$G$ is an auxiliary random variable introduced to make the coordinates of~$Y \sim \nu \mid G$ independent. Specifically, we first sample~$G$ uniformly from~$[t]$, and then set~$Y_i = 0$ for all~$i \ne G$, while~$Y_G$ is drawn uniformly from~$\{0,1\}$.
Equivalently, we can define the marginal distribution~$\nu$ over~$Y \in \{0,1\}^t$ as in our setting: with probability~$1/2$,~$Y = 0^t$, and with probability~$1/2$,~$Y = e_i$ for a uniformly chosen~$i \in [t]$. Then, define~$G$ as follows: if~$Y = e_i$, set~$G = i$; if~$Y = 0^t$, choose~$G$ uniformly in~$[t]$. Note that~$G$ is a randomized function of~$Y$, sampled \emph{after}~$Y$ is drawn, and is independent of the protocol transcript~$\Pi$ conditioned on~$Y$. Therefore, we have~$I(G; \Pi \mid Y) = 0$, and the chain rule gives:
\[
I(Y; \Pi) = I(Y, G; \Pi) = I(Y; \Pi \mid G) + I(G; \Pi) \geq I(Y; \Pi \mid G).
\]
Thus, a lower bound on the conditional information cost~$I(Y; \Pi \mid G)$ directly implies the same lower bound on the unconditional cost~$I(Y; \Pi)$ used here.
\end{remark}

For our proof, we need a slight variant of Lemma~\ref{lem:AND_lb}. Let~$p\leq \frac{1}{t}$ be some probability. 
We consider a distribution~$\mu=\mu_p$ in which every input bit~$Y_i$ is independently chosen to be~$1$ with probability~$p$ and otherwise~$0$.
We prove the following adaptation of Lemma~\ref{lem:AND_lb}. While it is straightforward to deduce it by repeating any of the previous proofs of Lemma~\ref{lem:AND_lb}, we give a proof that uses them in a black-box manner.

\begin{lemma}
    \label{lem:AND1}
    Suppose $\|\Pi_t(1^t)-\Pi_t(0^t)\|_1>0.1$, then $I_\mu(Y;\Pi)=\Omega(p)$.
\end{lemma}

\begin{proof}
We observe that~$\mu\left(0^t\right) = \left(1-p\right)^t \geq \left(1-\frac{1}{t}\right)^t \geq \frac{1}{4}$, and that for every~$i\in[t]$ also~$\mu\left(e_i\right) = p\cdot \left(1-p\right)^{t-1} \geq \frac{1}{4t}$.
Therefore, we may define a Boolean random variable $D$ such that: (1) $\Pr[D=1]=\Theta(p\cdot t)$; (2) $\mu|_{D=1} = \nu$. That is, conditioned on the event $D$ happening, $\mu$ becomes $\nu$. 
Note that~$D$ is independent of~$\Pi$.
Using Lemma~\ref{lem:AND_lb}, we get 
\begin{align*}
I_\mu(Y;\Pi) &=I_\mu(Y,D\ ;\ \Pi)-I_\mu(D;\Pi|Y)=I_\mu(Y,D \ ; \ \Pi)
=
I_\mu(D;\Pi) + I_\mu(Y;\Pi|D)\\
&=
I_\mu(Y;\Pi|D)\ge \Pr[D=1]\cdot I_\mu(Y;\Pi|D=1) = 
\Pr[D=1]\cdot  I_\nu(Y;\Pi) \\  &\ge 
\Theta(p\cdot t)\cdot\Omega(1/t) = \Omega(p). 
    \end{align*}

    \end{proof}

We denote the  problem from Lemma~\ref{lem:AND1} with this distribution $\mu$ by AND$_p^t$.

\subsection{Proof of Theorem~\ref{thm:EDISJ}}

For the remainder of the proof set $$p:=\frac{m}{2t |U|}.$$
Let $Y:=\{Y^j\}_{j=1}^{|U|}$ be $|U|$ instances of inputs to AND$^t$. Let $Y_i:=(Y^j_i)_{j=1}^{|U|}$ be the $i$-th player's input, which consists of the~$i$-th player's inputs in all of the~$|U|$ instances. 
Given a protocol $\Pi$ that solves EDISJ$_t$, we construct a protocol $\Pi'(Y_1,\ldots,Y_t)$ that we will then use to solve all of the instances of AND$^t$ defined above:

\begin{enumerate}
    \item 
    Each player $i\in [t]$ constructs an input set~$S_i$ as follows:
    \begin{enumerate}
        \item Set $S_i':=\{j\in U:~Y_i^j=1\}$;
        \item If $|S_i'|>\lceil\frac{m}{t}\rceil$, send `Fail', and the protocol fails; 
        \item Otherwise, set $S_i\subset U$ by adding $\lceil\frac{m}{t}\rceil-|S_i'|$ random elements from $U\setminus S_i'$ to $S_i'$.
    \end{enumerate}
    \item 
    Run $\Pi(S_1,\ldots,S_t)$ and return the message sent to the referee as the output~$\Pi'_t$.
\end{enumerate}

We first observe that if every instance~$Y^j$ is drawn from~$\mu_p$ then w.h.p no player will `Fail'.
\begin{lemma}
    When every~$Y^j$ is independently distributed as~$\mu_p$, then the probability that any~$S_i$ is larger than~$m/t$ is at most~$te^{-{m}/({2t})} < \frac{1}{4m\ln m}$.
\end{lemma}
\begin{proof}
    The size of~$S_i$ is distributed as~$\text{Binomial}(|U|, \frac{m}{2t|U|})$ and therefore a Chernoff bound gives that
    \[
    \Pr\left(|S_i| > \frac{m}{t}\right) \leq e^{-{m}/({2t})}
    .
    \]
    We then take a union bound over the~$t$ indices.
\end{proof}

Then, we also observe that the protocol's output~$\Pi'_t$ can be used to answer each of the AND$^t$ instances. 
\begin{lemma}\label{lem:Yj_answer}
    Let~$j\in U$ be a universe element. Let~$Y^j$ be an input to the~AND$^t$ problem. If for every~$u\neq j$ we draw~$Y^u$ independently from~$\mu_p$ and then run the above protocol~$\Pi'$ on~$Y_1,\ldots,Y_t$, then~$\Pi'_t$ implies the correct solution to AND$^t(Y^j)$ with probability $> 2/3 - o(1)$.
\end{lemma}
\begin{proof}
    Using~$\Pi'_t$, we may simulate the referee's part of the EDISJ$_t$ protocol with the choice of~$j$ as the exam question. This correctly computes AND$^t(Y^j)$ as long as no player mistakenly added~$j$ to its set in line (2) of the reduction, or if the reduction returned `Fail'. Nonetheless, the former happens with probability~$\leq p \cdot m = o(1)$ and the latter happens with probability~$\leq \frac{1}{4m\ln m}=o(1)$.
\end{proof}

For any $j\in U$, Lemma~\ref{lem:Yj_answer} implies that $\Pi'_t$ conditioned on $Y^j=0^t$ must be at statistical distance $>1/3-o(1)$ from $\Pi'_t$ conditioned on $Y^j=1^t$. Therefore, by Lemma~~\ref{lem:AND1} we 
have that when~$Y^j \sim \mu_p$ (like all other instances~$Y^u$) then
$$
I(Y^j;\Pi')=\Omega(p) = \Omega\left(\frac{m}{ t |U|}\right).
$$
Thus if every $Y^j\sim \mu_p$ independently, we also have 
$$
I(Y;\Pi')=\sum_{j\in U} I(Y^j;\Pi'|Y^{<j})=
\sum_{j\in U} I(Y^j \ ; \ \Pi', Y^{<j})\ge
\sum_{j\in U} I(Y^j;\Pi') =\Omega\left(\frac{m}{ t}\right).
$$

Let $F$ denote the Boolean random variable such that $F=1$ if $\Pi'$ fails (that is, any player outputs `Fail').
We observe that conditioned on~$F=0$, the distribution of~$X':=(S_1,\ldots,S_t)$, the input to the portocol~$\Pi$, is exactly~$\mu = \mu\left(t,m,U\right)$.
Since~$\Pi$ runs on the input~$X'$ when we do not fail, then~$I(Y; \Pi' \ | \ X', F=0) = 0$ as there is no additional information about~$Y$ seen by the protocol besides~$X'$.
We therefore have, 
\begin{align*}
I_\mu(\Pi ; X') &= I(\Pi' ; X' |F=0) =
I(\Pi'; X',Y |F=0) - I(\Pi'; Y |X' ; F=0)\\ &=
I(\Pi'; X',Y |F=0)  \ge 
I(\Pi'; Y |F=0) \ge I(\Pi';Y)-H(F) - \Pr[F=1]\cdot  I(\Pi'; Y |F=1)\\ 
&\ge \Omega\left(\frac{m}{ t}\right)-1 - t\exp\left(-\frac{m}{2t}\right)\cdot m \log|U| = \Omega\left(\frac{m}{ t}\right).
\end{align*}
The last inequality holds since even in the failure event, each player who transmits anything but `Fail' only holds at most $m/t$ elements, bounding the total entropy of their inputs by $m\log |U|$.

\hfill \qedsymbol{}

\section{Lower Bound}\label{sec:lower}
We begin with Section~\ref{subsec:EDISJtoF2}, in which we show how to reduce an instance of EDISJ to the problem of $F_2$ estimation on a stream. This reduction already gives a simple proof of Woodruff's~$\Omega(1/\varepsilon^2)$ lower bound~\cite{woodruff2004optimal} in the entire range of~$\varepsilon=\Omega(1/\sqrt{n})$.
Then, in Section~\ref{subsec:manyEDISJ} we show that we can pack~$\Theta\left(\log \left(\varepsilon^2 n\right)\right)$ instances of EDISJ into a \emph{single} stream of length~$n$. 
Solving each single instance would require~$\Omega\left(\frac{1}{\varepsilon^2}\right)$ space; We prove that solving \emph{all}~$\Theta\left(\log \left(\varepsilon^2 n\right)\right)$ instances requires~$\Omega\left(\log \left(\varepsilon^2 n\right)/\varepsilon^2\right)$ memory bits, as much as it would take to solve each of them independently.
Crucially, our instances are not going to be independent of each other, but to share certain stream elements.
Finally, we prove the following, which is the main theorem of this paper.
\begin{theorem}\label{thm:lowerbound}
    Let~$\mathcal{A}$ be a streaming algorithm that gives an~$\left(1\pm \varepsilon\right)$ approximation to the~$F_2$ of its input stream and succeeds with probability~$\geq 2/3$, for some~$\varepsilon=\Omega(1/\sqrt{n})$.
    Then, the space used by~$\mathcal{A}$ is~$\Omega\left(\log\left(\varepsilon^2 n\right)/\varepsilon^2\right)$.
\end{theorem}

\subsection{Reducing EDISJ to $F_2$}\label{subsec:EDISJtoF2}
Fix some~$n$,~$\varepsilon > \frac{2}{\sqrt{n}}$ and~$|U|>n^3$.
Denote by~$t:=\lfloor \varepsilon \sqrt{n} \rfloor \geq 2$, for simplicity we assume that~$t$ divides~$n$.
We reduce~EDISJ$_t$ to estimating~$F_2$ on a stream of length~$\left(1 \pm o\left(1\right)\right) n$ up to error~$\varepsilon$.
Let $X=(S_1,\ldots,S_t,x)$ be a valid input to~EDISJ$_t$, for simplicity we also assume that~$|S_i|=\frac{n}{t}$ for every~$i\in[t]$ (otherwise we may pad the sets).
We define a stream~$s(X)$ with elements from~$U$ that encodes~$X$ as follows: First, we write down the elements of~$S_1$ in arbitrary order, afterwards we write down the elements of~$S_2$, then of~$S_3$ and so on until those of~$S_t$; Finally, we write down~$x$ repeatedly~$k:=\lceil\frac{t}{\varepsilon}\rceil = \Theta\left(\sqrt{n}\right)$ times.
We note that the length of~$s(X)$ is~$\left(1+O\left(\frac{1}{\sqrt{n}}\right)\right)n$, and that the stream is naturally partitioned to~$t+1$ parts such that each part depends on the input of a single player (where the $(t+1)$-th `player' is the referee).

\begin{lemma}\label{lem:EDISJ_gap}
    A~$\left(1\pm \Theta\left(\varepsilon\right)\right)$-approximation to~$F_2\left(s\left(X\right)\right)$ implies the answer to EDISJ$_t(X)$.
\end{lemma}
\begin{proof}
    If EDISJ$_t(X)$=`No' then there are two possible cases: First, there is no intersection to the input sets, and thus~$x$ might appear in at most one of the sets; Hence,~$F_2\left(s\left(X\right)\right) \leq (n-1)\cdot 1^2 + 1\cdot \left(1+k\right)^2 = n + k^2 + 2k.$
    Second, there is a unique intersection to all input sets, but this intersection is not~$x$, who might still appear in at most one of the sets; Hence,~$F_2\left(s\left(X\right)\right) \leq (n-t-1)\cdot 1^2 + 1\cdot t^2 + 1\cdot (1+k)^2 = n+k^2+2k+t^2-t.$
    On the other hand, if EDISJ$_t(X)$=`Yes' then there is a unique intersection to all input sets who is also~$x$, and thus~$F_2\left(s\left(X\right)\right) = (n-t)\cdot 1^2 + (t+k)^2 = n + t^2 - t + k^2 +2tk.$
    Due to our choice of parameters,~$F_2\left(s\left(X\right)\right)=O(n)$ in all cases; Furthermore, the additive gap between the `Yes' case and the largest possible `No' case is~$$
    2tk - 2k = 2k(t-1) \geq kt \geq \frac{t^2}{\varepsilon} \geq \varepsilon n
    .$$
\end{proof}

Next, we observe that a streaming algorithm approximating~$F_2\left(s\left(X\right)\right)$ naturally implies a communication protocol for EDISJ$_t(X)$.
Let~$\mathcal{A}$ be a streaming algorithm that gives a~$\left(1\pm \Theta\left(\varepsilon\right)\right)$-approximation to~$F_2$, we define a communication protocol~$\mathcal{P}$ as follows.
The first player, who knows~$S_1$, can construct the part of~$s(X)$ corresponding to it and start running~$\mathcal{A}$ on that prefix of the stream. Denote by~$M_1$ the memory in~$\mathcal{A}$ at the end of that part. The first player communicates~$M_1$ to the second player, who has~$S_2$ and can thus construct the second part of~$s(X)$; as it also has the communicated~$M_1$ it can now continue the run of~$\mathcal{A}$ until the end of the second part of the stream and communicate the memory in its end,~$M_2$, to the next player. We continue similarly until the~$t$-th player communicates~$M_t$ to the referee, who has~$x$ and can thus construct the last part of~$s(X)$ and complete~$\mathcal{A}$'s run. 
The transcript of~$\mathcal{P}$ is~$\Pi := \left(M_1,\ldots,M_t\right)$, note that it depends only on~$X':=(S_1,\ldots,S_t)$ and not on~$x$, or equivalently, only on the first~$n$ elements of~$s(X)$.
Lemma~\ref{lem:EDISJ_gap} implies that if~$\mathcal{A}$ correctly approximates~$F_2$ with error~$\Theta(\varepsilon)$ then~$\mathcal{P}$ correctly solves EDISJ$_t$ with the same success probability.
We conclude using Theorem~\ref{thm:EDISJ} that over the distribution~$X' \sim \mu(t,n,U)$ we have~\[
    I\left(X' \ ; \ \Pi\right) = \Omega\left(\frac{n}{t}\right) = \Omega\left(\frac{\sqrt{n}}{\varepsilon}\right)
    .
    \]
\begin{corollary}
    For any~$\varepsilon=\Omega(1/\sqrt{n})$, approximating~$F_2$ up to multiplicative error~$(1\pm \varepsilon)$ requires~$\Omega(1/\varepsilon^2)$ bits of space.
\end{corollary}
\begin{proof}
    The length of~$\Pi = \left(M_1,\ldots, M_t\right)$ is at most~$t$ times the size~$M$ of the space used by the streaming algorithm~$\mathcal{A}$ in bits.
    Thus, 
    \[
    \Omega\left(\frac{\sqrt{n}}{\varepsilon}\right) \leq I\left(X' \ ; \ \Pi\right) \leq H(\Pi) \leq tM
    ,
    \] and therefore \[
    M \geq \Omega\left(\frac{\sqrt{n}}{\varepsilon t}\right) = \Omega\left(\frac{1}{\varepsilon^2}\right).
    \]
\end{proof}

For the consecutive parts of our lower bound, we need a slightly more flexible reduction from EDISJ to~$F_2$. In particular, we show that with a slight modification the above reduction yields the same lower bound even if we reduce from~EDISJ$_t$ for any~$2\leq t \leq \varepsilon \sqrt{n}$ rather than exactly~$t=\lfloor \varepsilon \sqrt{n}\rfloor$.
Fix any such~$t$. Denote by~$d := \lfloor \frac{\varepsilon^2 n}{t^2} \rfloor \geq 1$. For simplicity, we assume that~$d\cdot t$ divides~$n$.
We still use the universe~$U$ for the stream elements, but our~EDISJ$_t$ instances are defined over the larger universe~$U^d$ of~$d$-tuples of elements from~$U$.
We add the \emph{promise} that we only need to solve instances of EDISJ$_t$ in which no two distinct elements of~$U^d$ appearing in the input intersect each other. 
We consider instances of EDISJ$_t$ in which the set sizes are~$|S_i| = \frac{n}{dt}$.
For an instance~$X=\left(S_1,\ldots,S_t,x\right)$ the stream~$s(X)$ is constructed almost identically to the previous construction, except that when we write an element of~$U^d$ on the stream we write replace it with the~$d$ elements of~$U$ it consists of. 
We still repeat the referee's element~$x\in U^d$ at the end of the stream for~$k=\lceil\frac{t}{\varepsilon}\rceil$ times.
Thus, the size of the stream corresponding to~$X'$ is still~$t \cdot \frac{n}{dt} \cdot d = n$ and its elements are still elements from~$U$. The size of the suffix of the stream corresponding to the referee is~$k\cdot d = O\left(\frac{td}{\varepsilon}\right) = O\left(\frac{t\varepsilon^2 n}{\varepsilon t^2}\right)  = O\left(\frac{\varepsilon n}{t}\right)$ which is at most~$O(\varepsilon n)$.
We next observe that Lemma~\ref{lem:EDISJ_gap} still holds: 
In both the `Yes' and `No' cases, we have~$F_2\left(s\left(X\right)\right) \leq d \cdot \left(\frac{n}{d} + t^2 - t + k^2 +2tk\right) = O\left(n\right)$.
Furthermore, the gap between the `Yes' case and the largest possible `No' case is at least~$d \cdot \left(2tk - 2k\right) \geq \Omega(dtk) \geq \Omega\left(\varepsilon n\right)$.
We conclude as before that for~$\Pi = \left(M_1,\ldots,M_t\right)$ and input distribution~$X' \sim \mu\left(t,\frac{n}{d},U^d\right)$ we have~\[
    I\left(X' \ ; \ \Pi\right) = \Omega\left(\frac{n/d}{t}\right) = \Omega\left(\frac{t}{\varepsilon^2}\right)
    .
\]

\begin{remark}
    This modification implies that the~$\Omega(1/\varepsilon^2)$ lower bound follows even from the Exam version of the~$2$-party DISJ problem.
\end{remark}

\subsection{Packing many EDISJ instances into a single $F_2$ instance}\label{subsec:manyEDISJ}
We now construct a stream of length~$n$ that encodes within it many instances of~EDISJ.
Fix some~$n, \varepsilon > \frac{4}{\sqrt{n}}$ and~$|U|>n^3$.
For simplicity, assume that~$\varepsilon^-1$ is a power of two and that~$n$ is a power of four (otherwise we may round to the nearest ones); In particular, $\varepsilon \sqrt{n}$ is also a power of two and divides~$n$.

Let~$\ell\in [1, \log \left(\varepsilon \sqrt{n}\right)-2]$ be a \emph{level} index.
We construct a stream of length~$n$ that encodes an instance of~EDISJ$_{t}$ for~$t=2^{\ell}$, similar to that of Section~\ref{subsec:EDISJtoF2}. 
Denote by~$d:=\frac{\varepsilon^2 n}{4t^2}$.
Denote by~$X'=\left(S_1,\ldots,S_t\right),X=\left(X',x\right)$ a valid instance to~EDISJ$_{t}$ with set sizes~$|S_i| = \frac{n}{4dt}$ and universe~$U^d$.
As before, the only valid inputs are those in which all sets~$S_i\subset U^d$ are either disjoint or have a unique intersection, and furthermore every distinct elements of~$U^d$ appearing in the input do not intersect each other (i.e., each element of~$U$ is used in at most one element of~$U^d$ appearing in the input sets).

We start by partitioning the stream of length~$n$ into named parts.
We partition the stream into~$2^{\ell+2}$ consecutive blocks of size~$\frac{n}{2^{\ell+2}}$ each. 
We call each such block \emph{a bucket}; The buckets of indices divisible by four are called \emph{active buckets} and the buckets of other indices are called \emph{inactive buckets}, denote by~$B^{(\ell)}_1,B^{(\ell)}_2,\ldots,B^{(\ell)}_{2^{\ell}}$ the~$t$ active buckets, each of them is a consecutive subsequence of the stream of length~${n}/{2^{\ell+2}} = {n}/{4t}$.
Within each active bucket, we again partition the stream into consecutive blocks of size~$d$ each; We call each such sub-block \emph{a super-element}.
Every active bucket contains~$\frac{n}{2^{\ell+2} \cdot d} = \frac{n}{4dt}$ super-elements. 
We refer the reader to Figure~\ref{fig:several_instances} for an illustration of the active buckets in several levels. 

\subsubsection*{Reduction from DISJ$_t$:}
We follow the reduction of Section~\ref{subsec:EDISJtoF2} in our current terminology.
Given a valid input~$X$, we put the set~$S_i$ for $i\in [t]$ in place of the active bucket~$B^{(\ell)}_i$ in the stream, where each element~$x\in S_i \subset U^d$ fills an entire super-element within the active bucket.
Every element of the stream not within an active bucket is drawn uniformly and independently from~$U$.
We append to the described stream a suffix of length~$d\cdot k$, for~$k=\frac{t}{\varepsilon}$, which we fill with~$k$ repetitions of~$x$ written as a super-element each time.

We observe that with high probability approximating the~$F_2$ of the described stream up to error~$\Theta(\varepsilon)$ also solves EDISJ$_t(X)$: This is because with high probability (at least~$1-\frac{1}{n}$) the random elements of~$U$ placed in the non-active buckets are completely disjoint to any element appearing in~$X$, and thus subtracting~$\frac{3}{4} n$ from the~$F_2$ of the entire stream leaves us with the~$F_2$ of the sub-stream containing only the active buckets --- which is exactly the stream analyzed in Section~\ref{subsec:EDISJtoF2}.

Let~$\mathcal{A}$ be an algorithm that estimates the~$F_2$ of a stream up to~$\Theta(\varepsilon)$ error. 
Let~$0<j_1<j_2<\ldots<j_t<n$ be any~$t$ indices such that~$j_i$ is located between the end of the~$i$-th active bucket and before the beginning of the~$(i+1)$-th active bucket (or before the referee's suffix of the stream when~$i=t$).
Denote by~$M_{j_i}$ the state of $\mathcal{A}$'s memory after processing the~$j_i$-th stream element while running over the stream~$S=S(X)$ we get from the reduction.
Due to the stated above,~$\Pi=\left(M_{j_1},\ldots,M_{j_t}\right)$ is the transcript of a communication protocol that solves~EDISJ$_t(X)$.

Denote by~$\xi$ the distribution of a stream of length~$n$ where each element is drawn uniformly and independently from~$U$. 
We observe that when~$X' \sim \mu\left(t,\frac{n}{4d},U^d\right)$, the resulting stream without the referee's suffix~$S(X')$ is exactly distributed according to~$\xi$.
Thus, from Theorem~\ref{thm:EDISJ} we conclude the following.
\begin{corollary}\label{cor:one_level}
    For indices~$j_1<j_2<\ldots<j_t$ as above, we have~\[
    I\left( \ \left(B^{(\ell)}_1,B^{(\ell)}_2,\ldots,B^{(\ell)}_{t}\right) \ ; \ \left(M_{j_1},\ldots,M_{j_t}\right) \ \right) = \Omega\left(\frac{t}{\varepsilon^2}\right)
    ,
    \]
    where the memory states of the algorithm~$\mathcal{A}$ and the active buckets are with respect to a stream drawn from~$\xi$.
\end{corollary}

\subsubsection*{Information measure for a single level~$\ell$:}
Following the above, we define an information measure that encapsulates how much information about the active buckets of level~$\ell$ is given by the memory of the algorithm in an average index. 
From now on, we only consider the input stream distribution~$\xi$ for~$\mathcal{A}$, and denote the stream itself by~$X=\left(X_1,\ldots,X_n\right)\sim \xi$.

\begin{definition}
    For level~$\ell$, denote by
    \[
{I}_{\ell} := 
\sum_{j=1}^{n} I\left(X_{\left(j- \frac{n}{2^{\ell}}, \ j-\frac{n}{2^{\ell+1}}\right]} \ ;\ {M}_j \ \mid \ M_{\left(j- \frac{n}{2^{\ell}}\right)}\right)
    .
    \]
\end{definition}

We next give a lower bound for~$I_{\ell}$, based on Corollary~\ref{cor:one_level}.
\begin{lemma}\label{lemma:level_lb}
    $I_{\ell} > \Omega\left(\frac{n}{\varepsilon^2}\right)$.
\end{lemma}
\begin{proof}
    Let~$j_1$ be any index in the range~$\left(\frac{3}{4} \cdot \frac{n}{2^{\ell}}, \frac{n}{2^{\ell}}\right)$. 
    For~$1<i\leq 2^{\ell}$, denote by~$j_i := j_{i-1}+\frac{n}{2^{\ell}}$.
    We observe that each~$j_i$ is between the end of the~$i$-th active bucket~$B^{(\ell)}_i$ and the beginning of the next, so by Corollary~\ref{cor:one_level} we have~\[
    I\left( \ \left(B^{(\ell)}_1,B^{(\ell)}_2,\ldots,B^{(\ell)}_{2^{\ell}}\right) \ ; \ \left(M_{j_1},\ldots,M_{j_{2^\ell}}\right) \ \right) = \Omega\left(\frac{2^{\ell}}{\varepsilon^2}\right)
    .
    \]
    By applying Lemma~\ref{lem:streaminginfo} this also implies 
    \[
    \sum_{i=1}^{2^\ell} I\left(B_{i}^{(\ell)} \ ; \ M_{j_i} \ \mid \ M_{j_{(i-1)}}\right) > \Omega\left(2^\ell \cdot \frac{1}{\varepsilon^2}\right)
    ,\]
    where we abuse notation and treat~$M_{j_0}$ as an empty conditioning.
    Furthermore, each active bucket~$B_{i}^{(\ell)}$ is fully contained within~$X_{\left(j_i- \frac{n}{2^{\ell}}, \ j_i-\frac{n}{2^{\ell+1}}\right]}$; Hence,
    \[
    I\left(X_{\left(j_i- \frac{n}{2^{\ell}}, \ j_i-\frac{n}{2^{\ell+1}}\right]} \ ; \ M_{j_i} \ \mid \ M_{j_{i-1}}\right)
    \geq
    I\left(B_{i}^{(\ell)} \ ; \ M_{j_i} \ \mid \ M_{j_{i-1}}\right)
    .
    \]
    The sequences~$j_1,j_2,\ldots,j_{2^{\ell}}$ corresponding to each different~$j_1\in \left(\frac{3}{4} \cdot \frac{n}{2^{\ell}},\frac{n}{2^{\ell}}\right)$ are disjoint, and there are~$\Omega\left(\frac{n}{2^\ell}\right)$ of those. We thus conclude that \[
    I_\ell \geq \Omega\left(\frac{n}{2^\ell}\right) \cdot \Omega\left(2^\ell \frac{1}{\varepsilon^2}\right) = 
    \Omega\left(n \cdot \frac{1}{\varepsilon^2}\right)
    .
    \]
\end{proof}

\subsubsection*{Combining multiple levels:}
Crucially, while the analysis leading to the lower bound of~$I_\ell$ depended on the specific level~$\ell$, the input stream's distribution is the same across all levels.
Our final component is a type of direct-sum over all~$\Theta\left(\log\left(\varepsilon^2 n\right)\right)$ information measures~$I_\ell$.

\begin{lemma}\label{lem:index_partition}
    For any index~$j\in [n]$, we have \[
    I\left(X_1,X_2,\ldots,X_{j-1} \ ; \ {M}_j\right) \geq 
    \sum_{\ell=1}^{\log \left(\varepsilon \sqrt{n}\right)-2} I\left(X_{\left(j- \frac{n}{2^{\ell}}, \ j-\frac{n}{2^{\ell+1}}\right]} \ ;\ {M}_j \ \mid \ M_{\left(j- \frac{n}{2^{\ell}}\right)}\right)
    .\]
\end{lemma}
\begin{proof}
    The subsets~$X_{\left(j- \frac{n}{2^{\ell}}, \ j-\frac{n}{2^{\ell+1}}\right]}$ of~$X_1,\ldots,X_{j-1}$ are disjoint, and we may thus use the chain law of mutual information to get~\[
    I\left(X_1,X_2,\ldots,X_{j-1} \ ; \ {M}_j\right) \geq 
    \sum_{\ell=1}^{\log \left(\varepsilon \sqrt{n}\right)-2} I\left(X_{\left(j- \frac{n}{2^{\ell}}, \ j-\frac{n}{2^{\ell+1}}\right]} \ ;\ {M}_j \ \mid \ X_{\leq\left(j- \frac{n}{2^{\ell}}\right)}\right)
    .
    \]
    We conclude the proof by observing that
    \[
    I\left(X_{\left(j- \frac{n}{2^{\ell}}, \ j-\frac{n}{2^{\ell+1}}\right]} \ ;\ {M}_j \ \mid \ X_{\leq\left(j- \frac{n}{2^{\ell}}\right)}\right) =
    I\left(X_{\left(j- \frac{n}{2^{\ell}}, \ j-\frac{n}{2^{\ell+1}}\right]} \ ;\ {M}_j \ \mid \ M_{\left(j- \frac{n}{2^{\ell}}\right)}\right)
    ,
    \]
    as to both~$X_{\left(j- \frac{n}{2^{\ell}}, \ j-\frac{n}{2^{\ell+1}}\right]}$ and~${M}_j$ the conditioning on either~$X_{\leq\left(j- \frac{n}{2^{\ell}}\right)}$ or~$M_{\left(j- \frac{n}{2^{\ell}}\right)}$ is equivalent.
\end{proof}

We next define a global measure of information between the algorithm's memory and the inputs, then bound it using the single-level information measures.

\begin{definition}
    Denote by~\[
    \bar{I}:=\frac{1}{n} \sum_{j=1}^{n} I\left(X_{<j} \ ; \ M_j\right)
    .
    \]
\end{definition}

\begin{lemma}\label{lemma:all_levels_lb}
    $\bar{I} \geq \frac{1}{n} \sum_{\ell=1}^{\log \left(\varepsilon \sqrt{n}\right)-2} I_\ell$.
\end{lemma}
\begin{proof}
    Using Lemma~\ref{lem:index_partition} we have
    \begin{align*}
    \sum_{j=1}^{n}  I\left(X_{<j} \ ; \ {M}_j\right)
    &=
    \sum_{j=1}^{n} \sum_{\ell=1}^{\log \left(\varepsilon \sqrt{n}\right)-2} I\left(X_{\left(j- \frac{n}{2^{\ell}}, \ j-\frac{n}{2^{\ell+1}}\right]} \ ;\ {M}_j \ \mid \ M_{\left(j- \frac{n}{2^{\ell}}\right)}\right)\\
    &\geq
    \sum_{\ell=1}^{\log \left(\varepsilon \sqrt{n}\right)-2} \sum_{j=1}^{n} I\left(X_{\left(j- \frac{n}{2^{\ell}}, \ j-\frac{n}{2^{\ell+1}}\right]} \ ;\ {M}_j \ \mid \ M_{\left(j- \frac{n}{2^{\ell}}\right)}\right)\\
    &=
    \sum_{\ell=1}^{\log \left(\varepsilon \sqrt{n}\right)-2} I_\ell. 
    \end{align*}
\end{proof}

\begin{proof}[Proof of Theorem~\ref{thm:lowerbound}]
    By Lemma~\ref{lemma:all_levels_lb} and Lemma~\ref{lemma:level_lb} we have~\[
    \bar{I} \geq \sum_{\ell=1}^{\log \left(\varepsilon \sqrt{n}\right)-2} \frac{1}{n} I_\ell \geq \sum_{\ell=1}^{\log \left(\varepsilon \sqrt{n}\right)-2} \Omega\left(\frac{1}{\varepsilon^2}\right)
    \geq
    \Omega\left(\frac{\log\left(\varepsilon^2 n\right)}{\varepsilon^2}\right)
    .\]
    We conclude the proof by using Observation~\ref{obs:memoryfrominfo}.
\end{proof}

\section{Upper Bound}\label{sec:upper}

In this section we give a new tight upper bound on the problem of $F_2$ estimation. The upper bound is similar to the original AMS construction \cite{alon1996space}, with the modification that the random vectors onto which we calculate the projections have (random) disjoint support. 
To simplify the presentation we assume that the algorithm has access to public randomness -- similarly to prior works this assumption can be relaxed by replacing true randomness with $k$-wise independence for a constant $k$. The same algorithm with a different choice of parameters appeared in \cite{thorup2004tabulation} (see also \cite{charikar2002finding}), in the context of improving the update time of AMS. We include a full analysis here for completeness.

\begin{algorithm}[H]
    \caption{Partition-based $F_2$ algorithm}
    \label{alg:PartAlg}
    \begin{algorithmic}
        \State $P\gets \lceil 4/\varepsilon^2\rceil+1$;
        \State Let $H:U\rightarrow[P]$ and $\gamma:U\rightarrow\{-1,1\}$ be two random hash functions; 
        \State Initialize an array $A[1..P]$ of integer counters to $0$;
        \For{elements $x$ in the stream}
        \State $A[H(x)]\gets A[H(x)]+\gamma(x)$
        \EndFor
        \State
        Output $A=\sum_{i=1}^P A[i]^2$;
    \end{algorithmic}
\end{algorithm}

\subsection{Correctness analysis of Algorithm~\ref{alg:PartAlg}}

For the analysis, let $A$ be the random variable representing the algorithm's output, and $A_i$ the random variable representing the $i$-th cell of the array. 

Suppose the stream of length $n$ consists of elements 
$\{x_j\in U\}_{j=1}^n$ with frequencies $\{f_j\ge 0\}_{j=1}^n$\footnote{We allow some frequencies to be $0$ so that the counter on $j$ goes from $1$ to $n$}.

Let $\gamma_j:=\gamma(x_j)$, and let $B_{ji}\in\{0,1\}$ be the indicator random variable specifying whether $x_j$ got mapped to $A[i]$: $$B_{ji}:=\mathbf{1}_{H(x_j)=i}.$$

Then we have 
$$
A_i = \sum_{j=1}^n f_j \cdot B_{ji} \cdot \gamma_j;~~~A=\sum_{i=1}^P A_i^2
$$

Next, we compute $A$'s expectation and variance over the choices of the $H$ and $\gamma$ hash functions. 

Observe that for each $i$,
\begin{align*}
\E[A_i^2] &= \E\left[\sum_{j\in[n]} f_j^2\cdot B_{ji}^2\cdot \gamma_j^2 + 
\sum_{j_1\neq j_2\in[n]} f_{j_1} f_{j_2}\cdot B_{j_1i} B_{j_2i}\cdot \gamma_{j_1}\gamma_{j_2} \right]\\
&=\E\left[\sum_{j\in[n]} f_j^2\cdot B_{ji} \right] \\ &= \frac{1}{P} \cdot \sum_{j\in[n]} f_j^2
\end{align*}
And for each $i_1\neq i_2$:
\begin{align*}
\E[A_{i_1} \cdot A_{i_2}] &= \E\left[\sum_{j\in[n]} f_j^2\cdot B_{ji_1}\cdot B_{ji_2}\cdot \gamma_j^2 + 
\sum_{j_1\neq j_2\in[n]} f_{j_1} f_{j_2}\cdot B_{j_1i_1} B_{j_2i_2}\cdot \gamma_{j_1}\gamma_{j_2} \right]\\
&=0,\end{align*}
since for all $j$, $B_{ji_1}\cdot B_{ji_2}=0$.

Therefore, 
$$
E[A]= \sum_{i=1}^P E[A_i^2] = 
\sum_{j\in[n]} f_j^2 = F_2,
$$
and $A$ is an unbiased estimator of the $F_2$ moment. 

Next, we will calculate the variance $\var[A]$. Let $i_1\le i_2\le i_3\le i_4\in[P]$ be any four (not necessarily distinct) elements. We have
\begin{align*}
    \E[A_{i_1}\cdot A_{i_2}\cdot A_{i_3}\cdot A_{i_4} ] & =
    \sum_{j_1,j_2,j_3,j_4\in [n]} f_{j_1}f_{j_2}f_{j_3}f_{j_4}\cdot \E[B_{j_1i_1}B_{j_2i_2}B_{j_3i_3}B_{j_4i_4}]\cdot \E[\gamma_{j_1}\gamma_{j_2}\gamma_{j_3}\gamma_{j_4}]
\end{align*}
This equals to $0$ unless $i_1=i_2$ and $i_3=i_4$: for example, if $i_1<i_2\le i_3\le i_4$, then when $j_1\in\{j_2,j_3,j_4\}$ we have $B_{j_1i_1}B_{j_2i_2}B_{j_3i_3}B_{j_4i_4}=0$, and when  $j_1\notin\{j_2,j_3,j_4\}$, we have $\E[\gamma_{j_1}\gamma_{j_2}\gamma_{j_3}\gamma_{j_4}]=0$.

Therefore, we only need to consider terms of the form $\E[A_{i_1}^2\cdot A_{i_2}^2]$.

If $i_1<i_2$, we have 
\begin{align*}
    \E[A_{i_1}^2\cdot A_{i_2}^2] & =
    \sum_{j_1,j_2,j_3,j_4\in [n]} f_{j_1}f_{j_2}f_{j_3}f_{j_4}\cdot \E[B_{j_1i_1}B_{j_2i_1}B_{j_3i_2}B_{j_4i_2}]\cdot \E[\gamma_{j_1}\gamma_{j_2}\gamma_{j_3}\gamma_{j_4}]
    \\ &=
     \sum_{j_1\neq j_3\in [n]} f_{j_1}^2f_{j_3}^2\cdot \E[B_{j_1i_1}B_{j_3i_2}]\cdot \E[\gamma_{j_1}^2\gamma_{j_3}^2] \\ &=
     \sum_{j_1\neq j_3\in [n]} f_{j_1}^2f_{j_3}^2/P^2 \\&=
     \frac{1}{P^2}\cdot \left(F_2^2-\sum_j f_j^4\right)
     \\ &=    \frac{1}{P^2}\cdot \left(F_2^2-F_4\right)
\end{align*}
If $i_1=i_2$, we have
\begin{align*}
    \E[A_{i}^4] & =
    \sum_{j_1,j_2,j_3,j_4\in [n]} f_{j_1}f_{j_2}f_{j_3}f_{j_4}\cdot \E[B_{j_1i}B_{j_2i}B_{j_3i}B_{j_4i}]\cdot \E[\gamma_{j_1}\gamma_{j_2}\gamma_{j_3}\gamma_{j_4}]
    \\ &= 6\cdot \sum_{j_5<j_6\in [n]}
    f_{j_5}^2f_{j_6}^2 \cdot \E[B_{j_5i}B_{j_6i}]+
    \sum_{j\in [n]} f_j^4\cdot \E[B_{ji}]
     \\ &=\frac{3}{P^2}\cdot \left(F_2^2-\sum_j f_j^4\right) + \frac{1}{P}\cdot \sum_j f_j^4\\&=
     \frac{3}{P^2}\cdot \left(F_2^2-F_4\right) + \frac{1}{P}\cdot F_4
\end{align*}
We are now ready to compute $\var[A]$:
\begin{align*}
    \var[A]&= \E[A^2]-F_2^2 \\&=
    \sum_{i_1\neq i_2\in [P]} \E[A_{i_1}^2\cdot A_{i_2}^2] + \sum_{i\in [P]} \E[A_i^4] - F_2^2\\
    & = P\cdot (P-1) \cdot 
     \frac{1}{P^2}\cdot (F_2^2-F_4)+ 
     P\cdot \left( \frac{3}{P^2}\cdot \left(F_2^2-F_4\right) + \frac{1}{P}\cdot F_4\right)- F_2^2\\&=
     \frac{2}{P} \cdot F_2^2 - \frac{2}{P}\cdot F_4 
     \\ &<  \frac{2}{P} \cdot F_2^2 
\end{align*}

\begin{theorem}
    \label{thm:algmain}
    The expected relative $\ell_2$-error of Algorithm \ref{alg:PartAlg} is bounded by $\varepsilon$:
\begin{equation}
    \label{eq:thm1}
    \E\left[\left(\frac{A-F_2}{F_2}\right)^2\right]<\varepsilon^2
  \end{equation}
\end{theorem}
\begin{proof}
    Since $\E[A]=F_2$, we have 
   $$
        \E[(A-F_2)^2]  =\var[A] < \frac{2}{P} \cdot F_2^2 < \varepsilon^2 \cdot F_2^2,
   $$
   implying \eqref{eq:thm1}.
\end{proof}

\paragraph{Memory analysis of Algorithm~\ref{alg:PartAlg}}
In the word model, or if each element $A[i]$ of the array is stored using $O(\log n)$ bits, the memory cost of the algorithm is $O(P\log n)=O(\varepsilon^{-2}\cdot \log n)$ -- matching the memory cost of \cite{alon1996space}. We observe that in some regimes -- when $\log (\varepsilon^2 n)\ll \log n$ -- Algorithm~\ref{alg:PartAlg} requires asymptotically less memory, matching our lower bounds above.

\begin{claim} Suppose the stream length is $n\ge \Omega(\varepsilon^{-2})$. Then
    Algorithm~\ref{alg:PartAlg} can be implemented using $O(P\log (n/P))=O(\varepsilon^{-2}\cdot \log(\varepsilon^2 n))$ memory. 
\end{claim}

\begin{proof}
    Note that the state of the memory consists of $P$ integers satisfying 
    $\sum_i |A[i]| \le n$. Each integer $A[i]$ can be stored using a prefix-free code that only requires $2 \log(|A[i]|+1)+O(1)$ bits\footnote{In fact,  $\log|A[i]|+O(\log \log |A[i]|) $ suffices}.   

    The total memory cost of concatenating $P$ such prefix-free encodings is therefore  at bounded by 
    $$
    2\sum_{i\in[P]} \log(|A[i]|+1) + O(P) \le
    2 P\cdot  \log \left(\frac{\sum_{i\in[P]}|A[i]|}{P}+1\right) + O(P) = O(P \log(n/P)).
    $$
    Here the inequality follows from the concavity of the $\log$ function.
\end{proof}

\section{Summary and Open Problems}\label{sec:open}
After this work, the space complexity of~$F_p$ estimation is understood with nearly-tight bounds for all integral frequencies. 
Nonetheless, several natural questions remain open.

For non-integral moment estimation in the range~$p\in(1,2)$, there remains a gap between the previous lower bound of~$\Omega(1/\varepsilon^2+\log n)$ and upper bound of~$O(\log n / \varepsilon^2)$. This is the last remaining range of~$p$ for which the space complexity of frequency moment estimation is not well understood.

For second moment estimation, we prove a tight space bound for all~$\varepsilon=\Omega(1/\sqrt{n})$, which is~$\Theta(n)$ when~$\varepsilon=\Theta(1/\sqrt{n})$. By maintaining the complete frequency vector, an exact computation of the second moment is possible with~$O\left(n \log \left(\min\{n, 1+\frac{|U|}{n}\}\right)\right)$ space. A gap between~$\Theta(n\log n)$ and~$\Theta(n)$ space remains for the range of~$0<\varepsilon<1/\sqrt{\varepsilon}$. What is the exact space complexity of~$F_2$ estimation in the regime of very low error?

Finally, it reamins open whether~$F_2$-estimation can be improved in random-ordered streams and other models of non-adversarially ordered streams.

\subsection*{Acknowledgments}
The authors would like to deeply thank David Woodruff for several helpful conversations.

\noindent
MB's research is supported in part by the NSF Alan T. Waterman Award, Grant No. 1933331.

\noindent
OZ's research is supported in part by the Israel Science Foundation, Grant No. 1593/24.

\bibliography{main}
\bibliographystyle{alpha}

\end{document}